\documentclass[aps,twocolumn,showpacs,amsmath,amssymb,pra,superscriptaddress,floatfix,longbibliography]{revtex4-1}
\usepackage{braket}
\usepackage{amsmath}
\usepackage{amssymb}
\usepackage{mathtools}
\usepackage{graphicx}
\usepackage{dcolumn}
\usepackage{bm}
\usepackage{multirow}
\usepackage{leftidx}
\usepackage{color}
\usepackage{float}
\usepackage{xcolor}
\usepackage{soul}

\newcommand {\editms}[1]{\textcolor{blue}}
\newcommand {\editfr}[1]{\textcolor{green}}
\newcommand {\edittg}[1]{\textcolor{red}}

\usepackage{amsfonts}
\usepackage{eufrak}
\usepackage[german,english]{babel}

\begin{document}
\title{Simulated optical molasses cooling of trapped antihydrogen}
\author{Spencer J.~Walsh}
\affiliation{Department of Physics and Astronomy, Purdue University, West Lafayette,
Indiana 47907, USA}
\author{C. \O. Rasmussen}
\affiliation{CERN, Experimental Physics Department, CH-1211 Gen\`eve 23,
Switzerland}
\author{F.~Robicheaux}
\email{robichf@purdue.edu}
\affiliation{Department of Physics and Astronomy, Purdue University, West Lafayette,
Indiana 47907, USA}

\date{\today}

\begin{abstract}
We theoretically and computationally investigate the cooling of antihydrogen,
$\bar{\rm H}$, using optical molasses cooling. This updates the
results in Ref.~\cite{DFR2013} to the current capabilities of the
ALPHA experiment. Through Monte Carlo simulation, we show that $\bar{\rm H}$s
do not give the standard cooling even in an ideal optical molasses because
of their small mass and large transition frequency.
For optical molasses cooling in the ALPHA trap, the photons are constrained
to travel in one direction only. It is only through the phase space mixing
in the trap that cooling in all directions can be achieved. We explore the
nontrivial role that laser intensity plays in the cooling. We also investigate
the possibility for simultaneously cooling atoms in either of the
trapped ground states.

\end{abstract}

\maketitle

\section{Introduction}

The antimatter version of the hydrogen atom, $\bar{\rm H}$, is the
simplest atomic antimatter and as such offers several possibilities
for high precision comparison with hydrogen atoms.
Cold $\bar{\rm H}$ atoms were magnetically trapped 14 years
ago\cite{ALPHA2010} enabling
measurements of its properties.
Proposed and actual comparisons
included the 1S-2S transition frequency,\cite{ALPHA2017b,ALPHA2018}
the hyperfine ground state splitting,\cite{ALPHA2012,ALPHA2017}
charge,\cite{ALPHA2014,BCF2014,ALPHA2016}
and acceleration from gravity.\cite{HZR2014,ALPHA2023}
Other possible measurements (e.g. 1S-3S or 2S-$n$S or
2S-$n$P) of sufficient accuracy
would constrain the charge radius of the antiproton as has been
done for the proton.\cite{PAN2010}

The mechanism forming $\bar{\rm H}$, three body
recombination,\cite{RH12004,GO11991,MK11969}
leads to center of mass temperatures comparable to that of the
positron plasma. Since the magnetic trap only holds atoms with
less than $\sim 1/2$~K energy, the accuracy of measurements are
limited by the relatively high $\bar{\rm H}$ velocities. The need for colder
$\bar{\rm H}$ atoms led to several proposals for different types
of laser cooling and a successful implementation of an optical
molasses cooling based on the 1S-2P transition.\cite{ALPHA2021}

Whatever differences between normal hydrogen and $\bar{\rm H}$ atoms exist,
they will be small. Thus, different methods used to cool normal hydrogen
could serve as possible templates for cooling
$\bar{\rm H}$.\cite{SWL1993,CFK1996,ZVG2001,DK12006,WBP2011,MFM2014,MPM2015,GGG2018}
However, two important constraints eliminates many methods or limits
their effectiveness. The first is that collisional type cooling
(e.g. evaporative cooling or sympathetic cooling) is
unavailable due to the extremely low density ($<10^3$~cm$^{-3}$)
of $\bar{\rm H}$
in the trap and the annihilation of $\bar{\rm H}$ on normal matter.
The second constraint derives from
the geometry of the trap and the sources of the magnetic
fields reducing the effectiveness of laser cooling techniques. For
example, the coils that shape the magnetic fields limit the spatial
dependence of the magnetic field to relatively smooth variations.
Another example is the laser access leads to a small number of laser
beams most constrained to near the axis of the trap.

The most promising cooling technique, and the only one successfully
implemented,\cite{ALPHA2021}
is a simple optical molasses based on the 1S-2P transition.
The laser in this experiment is a nearly Fourier transform limited
pulsed laser with the pulse duration a couple 10's of nanoseconds
and the linewidth of a few 10's of MHz.
For this case, a single 121.6~nm beam, somewhat red detuned from the
transition in the magnetic field, cools the axial motion of the
$\bar{\rm H}$ atoms. For this geometry,
the random re-emission of the photon tends to
heat the radial motion. Because the magnetic trapping potentials are not
perfectly symmetric, the $\bar{\rm H}$ motion mixes the axial and
radial degrees of freedom\cite{SWS1994,ZFZ2018}
which can lead to cooling of all 3 directions.\cite{DFR2013,ALPHA2021}
The intensity of the laser has to be sufficiently low that there is time
for the mixing to take place between successive photon scattering
otherwise the atoms will heat on average.
This heating is strongest for smallest detuning where
the photon scattering rate is largest. The spatially varying magnetic
field also complicates the photoabsorption by giving substantial
changes in the detuning versus position in the trap.

There are three main updates to Ref.~\cite{DFR2013}. First, we
recognize that some of the changes from optimal detuning are due
to the small mass and large frequency of the cooling transition
(Sec.~\ref{sec:NoTrap}). Second, we use magnetic fields and
laser parameters more representative of the ALPHA experiment
(Sec.~\ref{sec:AT}) leading to lower simulated
temperatures. Lastly, we describe a possible method for
simultaneously laser cooling the 1Sc and 1Sd trapped populations
(Sec.~\ref{sec:simcool}) leading to an important improvement to
precision measurements.

In this manuscript, we revisit the simple optical molasses for hydrogen
atoms because the large frequency of the 1S-2P transition and small
atom mass leads to nonstandard results. We describe the Monte Carlo
simulation of the optical molasses cooling in the ALPHA trap.
We describe why cooling both trapped populations is not possible for
a single frequency but can be accomplished for 2 or more
frequencies. We give results that indicate the role played by detuning
and the intensity.

\section{Energy levels in magnetic field}\label{sec:ELB}

\begin{figure}
\resizebox{80mm}{!}{\includegraphics{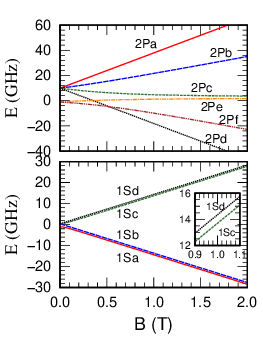}}
\caption{\label{fig:EvsB}
The splittings of the 1S and 2P states relative to their zero field
average. The hyperfine terms of the 2P states are too small to resolve.
The inset shows the 1Sd and 1Sc states near 1~T.}
\end{figure}

In the ALPHA experiment, $\bar{\rm H}$s are laser cooled in a magnetic trap
with a minimum $B\sim 1$~T\cite{ALPHA2021}.
This means the energy levels are strongly
changed from their low (or zero) field character, Fig.~\ref{fig:EvsB}.
Reference~\cite{RMR2017}
discusses the energy levels. The energies in a magnetic field are labeled
with a Roman letter at the end which, by convention, increases from low
to high energy for 1S states while increasing from high to low energy
for the 2P states (as ordered for small $B$).
There are two 1S states that can be trapped
called the 1Sc and 1Sd states. The 1Sd state has the positron and
antiproton spin aligned; in the 1Sc state, they are antiparallel.
To prevent losses from the photon emission step, the laser will excite the
2Pa state which has the positron spin and orbital angular momentum aligned
to give total positron angular momentum
$J=3/2,M=3/2$. Because of the different magnetic moments, the transition
frequency for the 1Sc-2Pa transition is approximately 675~MHz higher
than that for the 1Sd-2Pa transition at 1~T. The difference in frequencies
does not change much with increasing $B$; for example, at $B=$0.5~T,
the difference in frequencies is 660~MHz while the difference
in frequencies is 680~MHz at 1.5~T.

The hyperfine splitting of the 2P states is several 10's MHz compared to
the order 10 GHz splitting of the states. Thus, the antiproton spin is
nearly decoupled from the positron total angular momentum, $J$, at the
$\sim 1$~T of the ALPHA experiment.
The 1Sd-2Pa transition is exactly closed because all of the angular momenta
are aligned. The 1Sc-2Pa transition is not exactly closed because the 1S
state, with a more than $10\times$ larger hyperfine splitting than 2P,
has more mixing of the wrong direction antiproton
spin. After the 1Sc-2Pa transition,
the 2Pa state decays to the 1Sa state with a branching ratio from perturbation
theory of
\begin{equation}
{\cal B}=\left(\frac{1.42\; \rm GHz}{4\times 14.0\;{\rm (GHz/T)}\times B}
\right)^2
\end{equation}
where the numerator is from the 1S hyperfine splitting and the denominator
is from the Zeeman splitting when flipping the positron spin. For
$B\sim 1$~T, the branching ratio is $\sim 6.4\times 10^{-4}$.
Thus, the cooling of the 1Sc population in the ALPHA trap should use less than
$\sim 100$ photons to lose less than 10\% of the population to spin flip.
The $\bar{\rm H}$ can be cooled to steady state with less than $\sim 100$
photons so the spin flip is not a problem as long as the cooling is not
overextended.
For smaller $B$, the spin flip losses are larger and could become a serious
issue below $\sim 0.1$~T.

\section{Monte Carlo: no magnetic trap}\label{sec:NoTrap}

As discussed in the introduction, laser cooling of trapped $\bar{\rm H}$
is complicated by the changing transition frequency due to the motion
through the spatially varying $B$-field. The cooling is also
complicated by the fact that the photons only travel in one direction
and thus can only cool one component of the velocity. Cooling of all
directions relies on the
motion through the trap to mix the velocity components. To help with
understanding the changes to optical molasses cooling due to the
special circumstances of the ALPHA trap, we present in this section
results when the restrictions of the trap are removed.
Section~\ref{sec:EngShif} reminds the reader of a small energy shift
in the laser tuning that arises from the small mass and large
transition frequency for $\bar{\rm H}$.
Section~\ref{sec:IOM} describes an algorithm for laser cooling similar in
form to that used in the theory of the actual trap but idealizes the
physics by having fast mixing of the velocities and neglecting the
frequency shift with position; because of the small mass and large
transition frequency, this treatment does not lead to the standard
optical molasses temperature versus detuning.

\subsection{Energy shift}\label{sec:EngShif}

A well known recoil term is usually left out in descriptions of laser cooling,
due to its smallness in determining the energy
conservation of the transition. For $\bar{\rm H}$ cooling, this term is
large enough that it could cause some changes in the results. Consider the
energy before and after the photon absorption
\begin{equation}
E_g + \frac{1}{2}Mv^2 + \hbar\omega = E_e + \frac{1}{2}M|\vec{v}+\vec{v}_k|^2
\end{equation}
where $E_{e,g}$ are the internal excited,ground state energies, $M$ is
the mass of the atom, $\vec{v}$ is the atom velocity before the photon
absorption, $f=\omega /(2\pi)$ is the frequency of the photon,
and $\vec{v}_k=\hbar\vec{k}/M$
is the recoil velocity from the photon absorption with $k$ the photon
wave number. Solving for $\omega$
gives the resonance condition\cite{CDG1998}
\begin{equation}
\omega_r (\vec{v}) 
= \frac{E_e-E_g}{\hbar}+\vec{v}\cdot\vec{k}+\frac{\hbar k^2}{2M}
=\omega_0+\vec{v}\cdot\vec{k}+\frac{\hbar k^2}{2M}
\end{equation}
with $\omega_0/(2\pi)$ the nominal resonance frequency,
$\vec{v}\cdot\vec{k}$ is from the Doppler effect, and the last term is
a recoil contribution to the energy.

Typically, the last term is dropped because it is much smaller than the
linewidth or other interesting energy scales. For $\bar{\rm H}$, this
term corresponds to a frequency of 13.4~MHz, approximately 13\% of the
linewidth. In all that follows, we will consider this term to be added to
the $\omega_0$.

\subsection{Ideal Optical Molasses}\label{sec:IOM}

In this section, we describe a method for Monte Carlo simulations
of laser cooling of $\bar{\rm H}$ using an ideal optical
molasses\cite{CJF2005}. We use a simplified method that can easily be extended to
simulate cooling in the full trap, Sec.~\ref{sec:AT}, but does
not include the shift in detuning nor the changing velocity
as the $\bar{\rm H}$ moves through the trap but only
changes the direction of the velocity vector on
a time scale short compared to the times between the laser pulses.
Ideal optical molasses cooling assumes the photon has a wave vector, $\vec{k}$,
in a random direction relative to the atom's velocity, $\vec{v}$. The Doppler
effect leads to an effective detuning of
\begin{equation}\label{eq:detkv}
\Delta = \omega-\omega_r(\vec{v})=\Delta_0 - \vec{k}\cdot\vec{v}
\end{equation}
where $\Delta_0=2\pi\delta f$ with $\delta f$ the frequency detuning of the
laser in the lab frame including the shift described in the previous
section. A negative detuning leads to more absorption
when the atom's velocity is opposite to the photon propagation direction
resulting in a kick from the absorption step tending to slow the atom.
For the ideal optical molasses, we will take the photon emission
to be in a random direction which leads to heating on average. After
one absorption and emission event, the velocity changes by
\begin{equation}\label{eq:kick}
\vec{v}\to \vec{v} + \frac{\hbar\vec{k}}{M} - \frac{\hbar\vec{k}_{emit}}{M}.
\end{equation}
When $\hbar k/M$ is small compared to the atom speed, then the
ideal optical molasses leads to the lowest average atom kinetic energies
when $\Delta_0 = -\Gamma /2$ with $\Gamma$ the decay rate of the excited
state. For this detuning, the
minimum average energy is $\langle E\rangle_{\rm min}= 3\hbar\Gamma /4$
corresponding to $k_B T_{\rm min} = \hbar\Gamma /2$ where $k_B$ is
Boltzmann's constant\cite{CJF2005}. For general detuning,
the ideal optical molasses gives a steady state temperature\cite{CJF2005}
\begin{equation}\label{eq:Tvsdetu}
k_B T_{\rm om} = -\frac{\hbar\Gamma}{4}\left[ 1+ \left(\frac{2\Delta_0}
{\Gamma}\right)^2\right]\frac{\Gamma}{2\Delta_0}
\end{equation}
where we have assumed the Rabi frequency of the transition, $\Omega$, is
much smaller than $\Gamma$ so that power broadening can be ignored.

Our Monte Carlo simulation of the ideal optical molasses used the following
algorithm: (1) for each of $N$ atoms, randomize the direction of the atom's
velocity by keeping its speed but changing the direction by randomly
picking a point on the surface of a sphere
and for that atom compute the effective detuning, Eq.~(\ref{eq:detkv});
(2) compute the probability that the photon was absorbed using
$P = P_0/[1+(2\Delta /\Gamma )^2 ]$ with $P_0$ a small number of order 0.01;
(3) compute a random number;
if it is smaller than $P$, then a photon was absorbed;
(4) if a photon is absorbed, compute a random emission direction and use
Eq.~(\ref{eq:kick}) to update that atom's velocity. If we wanted to
simulate the effect of thermalization from elastic collisions between the
atoms, we added a step: (5) randomly pick pairs of atoms and rotate their
relative velocity to emulate a collision; repeated often enough step (5) leads to
a Maxwell-Boltzmann velocity distribution at temperature $k_B T = (2/3)
\langle KE\rangle $.

For $\bar{\rm H}$ cooled on the $1S-2P$ transition, $k \simeq 2 \pi /121.6$~nm,
$\Gamma \simeq 6.26\times 10^8$~s$^{-1}$, and $M$ is the hydrogen atom mass.
The recoil velocity $\hbar k/M\simeq 3.26$~m/s and the recoil energy
$\hbar^2k^2/2M\simeq 8.87\times 10^{-27}$~J$\simeq 0.642$~mK$k_B$. For an
ideal optical molasses, the expected lowest temperature is
$T_{\rm min} \simeq 2.39$~mK for detuning $\Delta_0=-\Gamma /2$.

\begin{figure}
\resizebox{80mm}{!}{\includegraphics{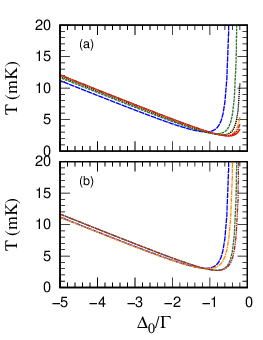}}
\caption{\label{fig:perfomp}
For an ideal optical molasses, the steady state temperature using the
Monte Carlo method versus the
lab frame detuning in units of $\Gamma$. For (a), there is no collision
between the atoms to achieve a Maxwell-Boltzmann velocity distribution.
The different lines are for different $\bar{\rm H}$ masses to illustrate
the role that the small atomic mass plays in the steady state
temperature: blue long dash
($1M$), green dash ($2M$), black dot ($4M$), orange dash-dot ($8M$),
maroon dash-dot-dot ($16M$), and
red solid is the ideal case, Eq.~(\ref{eq:Tvsdetu}). For (b), we
compare the effect that elastic collisions leading to thermalization has
on the steady state temperature: no collisions [blue long dash
($1M$), green dash ($2M$)], with collisions [orange dash-dot ($1M$),
maroon dash-dot-dot ($2M$)].
}
\end{figure}

The results of the simulations are shown in Fig.~\ref{fig:perfomp}.
Figure~\ref{fig:perfomp}(a) shows the results of the simulations when
we did not include collisions between the atoms, skipping step (5) of the
algorithm. This figure shows the results when the photon wavelength is
kept at 121.6~nm but we artificially change the mass of the $\bar{\rm H}$
from 1 to 16 times the actual mass by factors of 2 in order to illustrate
the role that the small atomic mass plays in the change of the steady
state temperature from the ideal result of Eq.~(\ref{eq:Tvsdetu}).
The blue long-dash
curve is for the actual $\bar{\rm H}$ mass. The red solid curve is the ideal
case, Eq.~(\ref{eq:Tvsdetu}).

The minimum temperature that can be reached for $1M$ happens
for a detuning of $\Delta_0\simeq -1.1\Gamma$ and gives
$T\simeq 3.05$~mK. The temperature for $1M$ at $\Delta_0=-\Gamma/2$ (the minimum
of Eq.~(\ref{eq:Tvsdetu})) is $\simeq 19$~mK
which is almost $8\times$ hotter than expected from the usual optical
molasses relations. From Fig.~\ref{fig:perfomp}(a), the higher masses
become progressively closer to the ideal optical molasses case, red solid line.
The reason for the discrepancy is the large size of the velocity kick,
3.26~m/s, compared to the thermal speed $\sqrt{k_B T/M}\sim 4.4$~m/s
at 2.39~mK (the minimum for the ideal optical molasses). This is similar to
the finding in Ref.~\cite{CWD1989} where the best detuning and lowest
temperature changes as the recoil energy becomes larger than the line
width. Interestingly,
the $1M$ temperature is less than the ideal case by $\sim 0.9$~mK
for larger magnitude detunings, $\Delta_0<-\Gamma$.

In the antihydrogen traps, the $\bar{\rm H}$ density is too low for
elastic collisions to play a role in thermalizing the distribution.
However,
in other atomic experiments of hydrogen, the density might be high enough
for elastic collisions to play a role. In Fig.~\ref{fig:perfomp}(b),
we show the results of laser cooling while including
elastic collisions between the atoms to
give a Maxwell-Boltzmann distribution. We show the results without
(blue long dash and green dash curves) and with (orange dash-dot and
maroon dash-dot-dot) elastic collisions. The collisions allow for
lower temperatures for $1M$ and $2M$. The difference
arises because the no collision case
leads to a few atoms with quite high kinetic energy. The collisions bring
them back to lower speeds where they can be more efficiently cooled.

\section{Monte Carlo: antihydrogen trap}\label{sec:AT}

The Monte Carlo simulation of cooling in the ALPHA trap is
more complicated than the algorithm in the previous section because the
atoms are confined to move in a magnetic trap. Our simulations are similar
to those in Ref.~\cite{DFR2013} but with details updated for the current
ALPHA trap. Also, see Ref.~\cite{RMR2017} for an overview for how the
$\bar{\rm H}$s are trapped and
probed.

Some of the features specific for the optical molasses cooling
include the path of the laser which is a straight line with
a $2.3^\circ$ tilt relative to the trap axis taken to be the
$z$-direction. Thus, the $\vec{k}$ is a constant.
The motion of the $\bar{\rm H}$ through the trap mixes the different
velocity components possibly allowing for three dimensional cooling.
Since this mixing is relatively slow, the detuning and intensity need
to be chosen to limit the heating of the radial motion due the photon
emission step.
The beam has a waist of 3.48~mm.
On the trap axis,
$B_\perp/B_z < 0.0003$ while at a
radius of 5~mm the $B_\perp/B_z < 0.05$ which means the direction
of $\vec{B}$ has little effect.
The laser is pulsed with a duration of 10's of
nanoseconds and with repetition rate of 50~Hz. The intensity is low enough
that much less than one photon is scattered from an $\bar{\rm H}$ per pulse
on average.
The laser cooling happens
over a time scale measured in hours which is possible due to the
cryogenic vacuum.

Because the atoms are in a large $B$-field, only one of the two trapped states
can be cooled if the laser frequency is held fixed. To give an idea,
the 1Sc-2Pa transition at $B=1$~T is approximately 675~MHz higher frequency
than the 1Sd-2Pa while the linewidth is approximately 100~MHz. If
the laser is set to cool the 1Sd $\bar{\rm H}$s, then there is almost no
photons scattered by atoms in the 1Sc state. While if the 1Sc $\bar{\rm H}$s are
cooled, the 1Sd $\bar{\rm H}$s will be heated because the laser will be
blue detuned relative to this transition. A possible method for
cooling both trapped states to the same extent is discussed in
Sec.~\ref{sec:simcool}.

Because the emission is from an $\ell =1,|m|=1$ state to an $\ell =0$
state, the probability for emission into different directions is not
uniform. If $\theta$ is the angle between the $B$-field direction
and the photon emission direction, the probability for photon emission
into a solid angle $d\phi d(\cos\theta )$ is proportional to
$1+\cos^2\theta$. This somewhat suppresses the emission perpendicular
to $\vec{B}$ relative to an isotropic distribution. Suppressing
perpendicular emission helps with
cooling in the ALPHA trap because
heating of the perpendicular directions due to photon emission is
one of the limiting factors.

\begin{figure}
\resizebox{80mm}{!}{\includegraphics{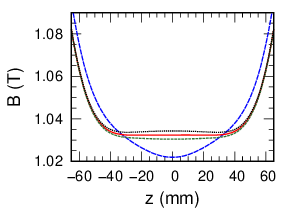}}
\caption{\label{fig:bfield}
The magnetic fields for the four simulations described in the text.
The red solid line is the ``flat" $B$-field, Sec.~\ref{sec:Bflat}
and the blue long dash line is the ``harmonic" $B$-field,
Sec.~\ref{sec:BHO}. The green dashed and black
dot curves are the slightly dipped and raised traps of
Secs.~\ref{sec:Bdipra}. The $z$ is the distance
from the trap center along the trap axis. The change in potential
energy for the $\bar{\rm H}$ motion
is approximately $2/3$~K/T, meaning 0.03~T corresponds to
20~mK~$k_B$; the change in detuning of the 1Sc or 1Sd to 2Pa
is approximately -14~MHz/T meaning 0.01~T change corresponds to
140~MHz redder detuning.
}
\end{figure}

One of the important features is that the detuning of the photon depends
on the position in the trap as well as the Doppler shift. The reason for
this is the $B$-field changes with position in the trap and the
$1S-2P$ resonance frequency increases with increasing $B$-field.
For negative detuning, the detuning becomes increasingly
negative with increasing $B$-field.
Figure~\ref{fig:bfield} shows the magnitude of the $B$-field on
the trap axis as a function of the position on the axis relative to the
trap center. There are 5 mirror coils and 2 solenoidal coils that
are used to shape the $B$-field on the axis. The red solid line
is a flat field similar to that used in the 1S2S spectroscopy measurements
in Ref.~\cite{ALPHA2018}. The blue long dash line gives a more harmonic
trap used in some of the simulations below. The green dashed and black
dot curves are the slightly dipped and raised traps of
Sec.~\ref{sec:Bdipra}.

The $B$-field scale can be converted to an energy scale by using
the magnetic moment for the 1S state which is approximately $2/3$~K/T.
This means that a 0.03~T change in $B$-field is a 20~mK~$k_B$ change in
potential energy. The change in the detuning is approximately -14~MHz/mT. Thus, a
0.03~T increase in $B$-field gives a 420~MHz more negative detuning.

The laser line width, estimated as 60~MHz full-width half-maximum (FWHM),
also modestly affects the results since the intrinsic linewidth is approximately
100~MHz. The laser line width is incorporated into the simulation by having
each pulse at a slightly different detuning randomly chosen from a
Gaussian distribution with a FWHM of 60~MHz. This leads to an effectively
broader transition although the effect is not large. This should be a good
approximation because the internal states are effectively a two level system
(the next allowed transition with similar transition matrix element
is detuned by approximately 300 linewidths),
the laser pulses are separated by sufficient time that
there are no atomic coherences (either in position or internal states)
between the pulses, the laser pulses are
nearly Fourier transform limited with a duration more than 10 times
longer than the lifetime, and the peak Rabi frequency is much less than
the 2P decay rate.
We verified that this
{\it is} a very good approximation (errors less than a couple percent)
by comparing to the results from numerically solving the
Optical Bloch equations for the same laser parameters.
In all of the figures and results, the linewidth, $\Gamma$,
is the natural radiative linewidth of the 2P state.

Reference~\cite{DFR2013} investigated the cooling from a hot distribution.
This requires a large amount of time to reach steady state because many of
the trajectories only rarely cross the laser beam. One can use a short
cut by starting with cold initial conditions and letting the atoms come to a
steady state. This allows a much faster determination of the steady state.
We launched the $\bar{\rm H}$s from a small region in the center of the trap
with a thermal velocity distribution so the average initial energy is similar
to the final, steady state energy. We ran the trajectories for either
1 or 3/2 simulation hours in the laser to ensure a steady state was achieved.

Because the laser is approximately on the axis, it is possible that the
laser cools the axial motion but heats up (or doesn't cool as well) the
radial motion. After the $\bar{\rm H}$s reach steady state, we compute
the averages $T_z\equiv 2\langle KE_z\rangle /k_B$ and $T_r\equiv 
\langle KE_z + KE_y\rangle /k_B$ while the laser is
on. We also calculate the average energy
of the $\bar{\rm H}$s. In the flat field, the average energy is approximately
$2k_BT$ for thermal distributions with $T<\sim 50$~mK.

\subsection{Cooling in a flat field}\label{sec:Bflat}

\begin{figure}
\resizebox{80mm}{!}{\includegraphics{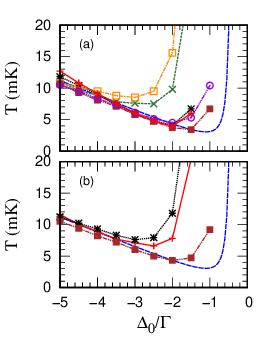}}
\caption{\label{fig:TvsDcomp}
Comparison of steady state temperatures versus detuning for different
pulse energy. The calculations were done for the flat field of
Fig.~\ref{fig:bfield}. In both (a) and (b), the blue long dash is the same as
the $1M$ in Fig.~\ref{fig:perfomp}. For (a) the curves are:
red solid ($+$) $T_z$ for 1 nJ, green dash ($\times$) $T_r$ for 1 nJ,
black dot ($*$) $T_z$ for 2 nJ, orange dash-dot (open square) $T_r$ for 2 nJ,
maroon dash-dot-dot (solid square) $T_z$ for 2 nJ with artificial mixing,
and purple dash-dot-dot-dot (open circle) $T_r$ for 2 nJ with artificial mixing.
For (b) the curves are: red solid ($+$) $T=(2T_r+T_z)/3$ for 1 nJ,
black dot ($*$) $T$ for 2 nJ, and
maroon dash-dot-dot (solid square) $T$ for 2 nJ with artificial mixing.
}
\end{figure}

Figure~\ref{fig:TvsDcomp} compares the steady state temperatures for
two different laser energies per pulse: 1~nJ or 2~nJ. These calculations
were done for the flat field in Fig.~\ref{fig:bfield}. For comparison,
the ideal case for $1M$ from Fig.~\ref{fig:EvsB} is included as the blue long dash
line. Both the $T_z$ and $T_r$ are shown in (a) while the combination
$T = (2 T_r+T_z)/3$ is shown in (b).
At larger detuning, all of the temperatures are approximately that for
the ideal optical molasses for $1M$. However at smaller detuning, the
temperatures in the trap tend to be higher than the ideal case with
$T_r>T_z$. Also, the $T_r$ is larger for the 2~nJ pulses than for the
1~nJ pulses. As the magnitude of the detuning decreases, this effect
becomes larger and is notable for $|\Delta_0|<3\Gamma$.
This indicates that the coordinate mixing by the $\bar{\rm H}$
motion through the trap is too slow to keep up with the cooling in $z$ and
heating in $r$. The rate for scattering photons
increases as the $|\Delta_0|$ decreases. Roughly speaking, the
mixing in the trap can keep up with the cooling of $z$ and heating in $r$
for $|\Delta_0| > 3.5\Gamma$ but can't for smaller detuning. 
Also, the mixing rate decreases as the axial motion is cooled which
can lead to radial heating if the axial cooling is too fast.
Because
more photons are scattered with 2~nJ pulses, the effect is larger for
the larger intensity.

For a clearer picture of this effect, we repeated the simulation but
added artificial mixing into the coordinates. This was done by randomly
making small changes to the direction of the $\bar{\rm H}$ velocity.
We chose parameters so that $\langle \vec{v}(t)\cdot\vec{v}(0)\rangle 
=v^2(0)e^{-t/T}$
with $T=50$~s when there are no trapping forces. These are the artificial
mixing curves in Fig.~\ref{fig:TvsDcomp} which are at much lower
temperature and more closely track the ideal molasses case at
small $|\Delta_0|$.
In Ref.~\cite{DFR2013}, the lowest average energy in Table 2 was
32~mK which corresponds to a temperature of $\sim 16$~mK. This result
is substantially hotter than Fig.~\ref{fig:TvsDcomp} because substantially
higher energy laser pulses were used: 50~nJ with 10~Hz repetition rate
and circular polarization.
The current simulations with 2~nJ at 50~Hz repetition rate and linear
polarization is effectively 10$\times$ fewer photons per second. This is also why
the simulations in Ref.~\cite{DFR2013} were for only a couple 100 seconds
while the current simulations are for 1000's of seconds.

\subsection{Cooling in a slightly dipped or raised field}\label{sec:Bdipra}

We simulated cooling when the middle mirror coil is used to generate a
dip in the magnetic trap, Fig.~\ref{fig:bfield} green dash curve.
This leads to a clear minimum of
$\sim 2$~mT compared to the flat field. On an energy scale,
this corresponds to $\sim 4/3$~mK~$k_B$ which is much smaller than the
scale in Fig.~\ref{fig:TvsDcomp}. We also simulated a small hump
by changing the current in the middle mirror,
Fig.~\ref{fig:bfield} black dot curve, with a clear maximum
of $\sim 2$~mT compared to the flat field. Because these are
small changes in energy compared to the temperatures in
Fig.~\ref{fig:TvsDcomp}, it might be expected that the
results will not be much different from the previous section.

Similar to a result in Ref.~\cite{DFR2013}, we found that the $\bar{\rm H}$
coldest temperature was a bit higher than in the flat trap for the slightly
dipped field.
The $\bar{\rm H}$ coldest temperature was approximately 10\% higher than
in the previous section. We also found the $\bar{\rm H}$
coldest temperature was a bit lower than in the flat trap for the slightly
raised field. In these two cases, the $T_z$ was hardly changed from 
that for the flat field. The changes were in the $T_r$ and reflect
the longer or shorter mixing times of the trajectories
in the dipped or raised field.

\subsection{Cooling in a nearly harmonic trap}\label{sec:BHO}

The harmonic trap, blue long dash curve in Fig.~\ref{fig:bfield}, has
only a tiny region where the detuning has smallest magnitude. This
suggests that the problems with the photon scattering rate being
higher than the coordinate mixing rate will be less. The actual case
is that the harmonic trap has much smaller mixing rate between the
coordinates. We find that the temperature differences are much more
extreme than the previous
section. For example, at $\Delta_0=-3\Gamma$ and 2~nJ, $T_z\simeq 6$~mK and
$T_r\simeq 24$~mK for the harmonic trap while the flat trap has $\simeq 6$
and $\simeq 9$~mK. Thus, nearly harmonic traps appear to be a poor choice
for laser cooling of only the axial motion. Similarly, making a much larger
hump in the previous section also leads to less effective
cooling because the two side
wells become more harmonic like.

\section{Simultaneous cooling 1Sc and 1Sd}\label{sec:simcool}

Because magnetic fields in the ALPHA experiment slowly drift with time,
measurements that require both trapped states (e.g. Ref.~\cite{ALPHA2017})
requires simultaneous measurements of the 1Sc and 1Sd
trapped populations for high
accuracy. This is straightforward when measurements are
performed on uncooled populations. There are no mechanisms to distinguish
the formation of antiproton up versus down states which is the main
difference between 1Sc and 1Sd.

Unfortunately, problems arise when laser cooling both populations and trying
to achieve the same distributions for each. The 1Sc-2Pa transition frequency
is 675~MHz larger at $B\sim 1$~T than the 1Sd-2Pa. Detuning to
-250~MHz for the 1Sd-2Pa transition gives -925~MHz detuning for the
1Sc-2Pa transition. Thus, the cooling is extremely slow for the
1Sc states and the final distribution is at a substantially larger
temperature with, essentially, uncooled $\bar{\rm H}$s.\cite{ALPHA2021}
In principle, the 1Sc and 1Sd states could be mixed using microwaves
with frequencies less than 1~GHz, but such long wavelength light
does not propagate down the $\sim 22$~mm radius ALPHA trap; this
would require a special resonator and microwave input in the next
generation trap.

One possibility that would lead to nearly the same distributions would use
(say) -250~MHz detuned for the 1Sd-2Pa transition (-925~MHz detuning for
1Sc-2Pa), then 675~MHz higher
than this (425~MHz detuning for 1Sd and -250~MHz detuning for 1Sc), and then
675~MHz higher (1100~MHz detuning for 1Sd and 425~MHz detuning for 1Sc).
If these frequencies are interleaved, the very far off resonance photons
(1100~MHz for 1Sd and -925~MHz for 1Sc)
would contribute almost no cooling or heating. The idea is to use only
the lowest frequency until the 1Sd $\bar{\rm H}$s
reached a steady state. Then the laser
pulses would alternate between $f$ or $f+675$~MHz. After the
1Sc are also cooled, then the frequencies would cycle
$f$ then $f+675$ then $f+1350$~MHz (repeat) so both populations experience
the same cooling. This method would work if
interleaving (say) -250~MHz detuning with 425~MHz detuning leads
to steady state cold atoms in the 1Sd state.

\begin{figure}
\resizebox{80mm}{!}{\includegraphics{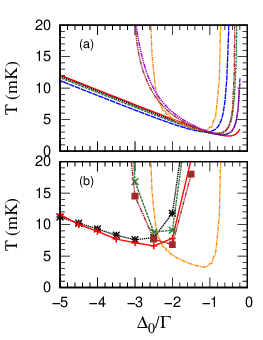}}
\caption{\label{fig:TvsDcomp2f}
(a) is similar to Fig.~\ref{fig:perfomp}(a). Four curves are from this
figure where the ideal one frequency optical molasses cools $\bar{\rm H}$s
with different masses to a steady state temperature: blue long dash
($1M$), green dash ($2M$), black dot ($4M$), and
red solid is the ideal case, Eq.~(\ref{eq:Tvsdetu}).
Three of the curves use the two frequency cooling where every other
photon pulse has detuning $\Delta_0$ or $\Delta_0+2\pi 675$~MHz.
The curves are: orange dash-dot ($1M$),
maroon dash-dot-dot ($2M$), purple dash-dot-dot-dot ($4M$).
(b) The orange dash-dot is the same from
(a) while the black dot ($*$) is $T=(2T_r+T_z)/3$
from the constant full simulation in the ALPHA trap with 2~nJ pulses,
Fig.~\ref{fig:TvsDcomp}(b). The green dash ($\times$) is the two
frequency results for 2~nJ and the maroon dash-dot-dot (solid square)
is the two frequeqncy results for 1~nJ.
}
\end{figure}

Before treating the cooling in the ALPHA trap, we examine the stability of the
ideal optical molasses with two frequencies.
Figure~\ref{fig:TvsDcomp2f}(a) shows the steady state temperature
for ideal optical molasses for 3 different $\bar{\rm H}$ masses:
$1M$, $2M$, and $4M$. Three of the curves are when there is only
one frequency and are the same as Fig.~\ref{fig:perfomp}(a):
blue long dash ($1M$), green dash ($2M$), and black dot ($4M$). The
red curve is the ideal optical molasses, Eq.~(\ref{eq:Tvsdetu}).
Three of the curves are the steady state temperature when two
frequencies are used: orange dash-dot ($1M$), maroon dash-dot-dot
($2M$), and purple dash-dot-dot-dot ($4M$).
We first ran at detuning $\Delta_0$ until
a steady state was reached. Then the laser pulses alternated between
$\Delta_0$ and $\Delta_0+2\pi 675$~MHz.
As expected, alternating with a blue detuned pulse leads to hotter
steady states. The $2M$ and $4M$ cases behave more as we were
expecting: the smaller $|\Delta_0|$ being less affected by the blue
detuned photons.
The $1M$ case does not behave in quite the
same way. Probably this is because smaller $|\Delta_0|$ leads to
stronger heating for $1M$ (see the blue long dash curve).
The hotter $\bar{\rm H}$s interact more strongly with the blue detuned
pulses which increases their temperature further.
Despite this, the orange dash-dot curve has a region
$-2\Gamma <\Delta_0 <-\Gamma$ of nearly unchanged cooling. Comparing this region
to the steady state temperatures in Fig.~\ref{fig:TvsDcomp}(b),
suggests that two frequency cooling might work in the ALPHA trap.

The results from simulations in the trap are shown in
Fig.~\ref{fig:TvsDcomp2f}(b) as the green dash ($\times$) for 2~nJ pulses
and maroon dash-dot-dot-dot (solid
square) for 1~nJ pulses. There appears to be a small range of detuning
(roughly $-3\Gamma < \Delta_0 < -2\Gamma$)
that leads to decent cooling even with the heating from the blue
detuned frequency. This range is narrow because the cooling in the
trap without the blue detuned pulses is not good for $-2\Gamma < \Delta_0$
and the blue pulses lead to heating for $\Delta_0 < -3\Gamma$.
The two detunings $-2.5$ and $-2\Gamma$ have nearly the same temperature
as the one frequency case (these are the black dot ($*$) and
red solid ($+$) curves).
The simulations are somewhat unrealistic in that the frequency was held
constant for 1 second before switching to a different frequency. In
practice, the frequencies probably can not be switched that often.
However, it is only necessary to change the frequencies on a time scale
where a small number of photons are scattered before the change.
Depending on the trap and laser details, probably this would be on
the several minute scale. These simulations suggest that the
1Sc and 1Sd states can be simultaneously cooled in the ALPHA trap.

\section{Summary}

Laser cooling
of $\bar{\rm H}$s has been demonstrated in Ref.~\cite{ALPHA2021}.
We have updated the laser cooling results of Ref.~\cite{DFR2013} to
better reflect the parameters of the ALPHA experiment. The update
includes laser parameters and magnetic field simulation.

Simulations show that even ideal optical molasses is complicated for
$\bar{\rm H}$s. For ideal optical molasses, the optimum detuning is
$\Delta_0 = -\Gamma /2$ leading to a minimum temperature of
$k_BT=\hbar\Gamma /2$ where $\Gamma$ is the decay rate of the upper state.
Because $\bar{\rm H}$s are cooled on the $1S-2P$ transition, the small
mass and large photon energy leads to relatively large recoil velocity
and energy. For untrapped $\bar{\rm H}$s, we find that the optimum detuning is
$\Delta_0\simeq -1.1\Gamma$ with a minimum temperature of
$k_BT\simeq 1.28\hbar\Gamma /2$; a detuning $\Delta_0 = -\Gamma /2$
leads to a temperature $k_BT\sim 8\times \hbar\Gamma /2$.

The spatially varying magnetic fields of the ALPHA trap adds a complication
to the cooling due to the constrained geometry for the laser. Three
dimensional cooling relies on the atom motion to scramble the
velocity vector. For perfectly separable motion in $x,y,z$, the
atoms could not be cooled. Larger energy per laser pulse leads to
a higher photon scattering rate which means the atoms' motion along
the laser can be cooled faster.
However, the limits of velocity mixing means that larger energy per
pulse can lead to higher steady state temperature. Larger energy per
laser pulse requires larger magnitude detuning to reach lower temperatures.
For the cases simulated, the lowest temperatures were achieved with a
detuning $\Delta_0\sim -2.5\Gamma$. We also simulated different shapes
for the magnetic trap which can affect the final temperature.

We also simulated the possibility for simultaneously cooling the
1Sd and 1Sc populations. The large magnetic field of the ALPHA trap
leads to  a 675~MHz detuning between the transitions starting in
the 1Sd and the 1Sc states. By interleaving three frequencies in the
laser pulses, we found a small region of detuning where both populations
can be cooled to the $\sim 10$~mK regime.

Although we have not simulated the following modifications, the outcomes
seem to follow from the results presented above. One possible modification
would be to increase the angle of the laser with respect to the trap
axis. While this is not possible in the current ALPHA experiment,
a larger angle (say between 20-40$^\circ$) would directly cool two
of the spatial coordinates (for example, $x$ and $z$) and only require
mixing of the coordinate perpendicular to the plane defined by the
laser and the trap axis. Another possibility is to
use stronger laser pulses with larger detuning early in the cooling
cycle and then transition to weaker pulses and smaller detuning near
the end of cooling; this should lead to faster cooling at early times
and lower temperatures at later times.


Data plotted in the figures is available at~\cite{data}.

\begin{acknowledgments}
FR thanks Claudio Lenz Cesar and AbdAlGhaffar K. Amer for insightful
conversations. A portion of the manuscript was written while FR was supported
at ITAMP.
This work was supported by the U.S. National Science Foundation under Grant
No. 2409162-PHY and
through a
grant for ITAMP at Harvard University.
The calculations were performed on a cluster run by
the Department of Physics and Astronomy, Purdue University.
\end{acknowledgments}

\bibliography{LaserCoolH}

\end{document}